\documentclass[aps,superscriptaddress,nofootinbib,preprintnumbers,showpacs,twocolumn,prd]{revtex4}

\usepackage{txfonts}

\usepackage{amssymb}

\usepackage{indentfirst}

\usepackage{bm}% bold math

\usepackage{epsfig}

\usepackage{epsf}

\usepackage{graphicx}

\newcommand{\be}{\begin{equation}}
\newcommand{\ee}{\end{equation}}
\newcommand{\ba}{\begin{eqnarray}}
\newcommand{\ea}{\end{eqnarray}}

\newcommand{\kp}{\boldsymbol{k}_T}

\newcommand{\p}{\perp}

\begin{document}

\title{Updating Boer-Mulders functions from unpolarized $pd$ and $pp$ Drell-Yan data }

\author{Zhun Lu}
\affiliation{Department of Physics, Southeast University, Nanjing
211189, China}
\affiliation{Departamento de F\'\i sica y Centro de Estudios
Subat\'omicos, Universidad T\'ecnica Federico Santa Mar\'\i a,
Casilla 110-V, Valpara\'\i so, Chile}

\author{Ivan Schmidt}
\affiliation{Departamento de F\'\i sica y Centro de Estudios
Subat\'omicos, Universidad T\'ecnica Federico Santa Mar\'\i a,
Casilla 110-V, Valpara\'\i so, Chile}

\begin{abstract}\
We extract the Boer-Mulders functions for the proton by combining
the unpolarized $p \,d$ and $p p$ Drell-Yan data measured by the
E866/NuSea Collaboration by the assumption that the $\cos 2\phi$
asymmetry is from the Boer-Mulders function. Using the extracted
Boer-Mulders functions, we present the predictions for the $\cos 2
\phi$ asymmetries in future $pp$ experiments at J-PARC and $p
\bar{p}$ experiments at PANDA and PAX.

\end{abstract}

\pacs{13.85.Qk, 14.20.Dh}

\maketitle

\section{introduction}

In a recent work~\cite{blms08a} the Boer-Mulders functions for the
proton have been extracted from the $\cos 2\phi$ angular asymmetry
measured in the unpolarized $p\,d$ Drell-Yan process by the
E866/NuSea Collaboration~\cite{e866} at FNAL. Much earlier, the
first $\cos 2 \phi$ asymmetries of dilepton production had been
measured two decades ago by the NA10~\cite{na10} Collaboration and
E165~\cite{conway} Collaboration, but for the $\pi$-nucleus
Drell-Yan processes, showing that the magnitude of the asymmetries
is around 30\% at most. This substantial $\cos 2 \phi$ angular
dependence that violates the so-called Lam-Tung relation~\cite{lt78}
predicted by perturbative QCD, belongs to the remaining challenges
which need to be understood from QCD dynamics. Several attempts have
been made to interpret these data, including QCD vacuum
effects~\cite{bnm93,bbnu} and higher-twist
mechanisms~\cite{bbkd94,ehvv94}. Furthermore, in the last decade
significant efforts have been put forward on the understanding of
the $\cos 2 \phi$ angular dependence from the view point of the
transverse momentum dependent (TMD) Boer-Mulders function
$h_1^{\p}(x,\boldsymbol{p}_T^2)$~\cite{bm}. It was shown~\cite{boer}
by Boer that the angular dependence can be related to the product of
two functions $h_1^{\p}$, each of which comes from one of the
incident hadrons. The Boer-Mulders function describes a correlation
between the transverse spin and the transverse momentum of a quark
inside an unpolarized hadron. Despite the naive T-odd nature of the
Boer-Mulders functions and of their chiral-even partner the Sivers
functions~\cite{sivers}, it has been shown that they can originate
from inital/final state interactions~\cite{bhs02,bbh03} between the
struck quark and the spectator of the nucleon, which are important
to ensure the gauge invariance of the TMD distribution
functions~\cite{collins02,belitsky,bmp03}. Recently a lot of
theoretical studies and phenomenological
analysis~\cite{gg02,pobylista,yuan,bsy04,lm04,
radici05,sissakian05,sissakian06,Lu:2006ew,Barone:2006ws,Lu:2007kj,
Burkardt:2007xm,gamberg07,lms07,gamberg072,Barone:2008tn,
Bacchetta:2008af,Wakamatsu:2009fn,Courtoy:2009pc,Gamberg:2009uk} on
Boer-Mulders functions have been carried out.

The first measurement of the $\cos2\phi$ asymmetry in the
nucleon-nucleon interacting Drell-Yan process by the E866/NuSea
Collaboration makes the attempt on extracting the proton
Boer-Mulders function possible. In Ref.~\cite{blms08a} we
parameterized the Boer-Mulders functions for $u$, $d$, $\bar{u}$ and
$\bar{d}$ quarks, which were fitted to the E866/NuSea $p\,d$ data,
based on the assumption that the $\cos2\phi$ asymmetry comes only
from the Boer-Mulders effect in the region $q_T^2 \ll Q^2$, where
$q_T$ and $Q$ are the transverse momentum and the invariant mass of
the lepton pair. In our parametrization the transverse momentum
dependence of $h_1^\p(x,\boldsymbol{p}_T^2)$ was modeled by the
Gaussian ansatz:
\begin{equation}
h_1^{\p \,q}(x,\boldsymbol{p}_T^2)=h_1^{\p \,q}(x)\,\frac{1}{\pi
p_{bm}^2}\, \exp\left(-\frac{\boldsymbol{p}_T^2}{p_{bm}^2}\right).
\end{equation}
The parameter $p_{bm}^2 $ describes the Gaussian width of the
transverse momentum distribution. The $x$-dependence of $h_1^{\p
q}(x)$ was further parameterized to be proportional to the
unpolarized parton distribution $h_1^{\p \,
q}(x)=x^c\,(1-x)\,f_1^{\,q}(x)$, where the parameter $c$ is the same
for all flavors. The large-$x$ behavior of $h_1^{\p \,q}(x)$
compared with $f_1^{\,q}(x)$ was taken into account by the
factor$(1-x)$ following the argument in Ref.~\cite{bro06}.

More recently the E866/NuSea Collaboration reported new
measurements~\cite{e866pp} on the $\cos 2\phi$ asymmetry in the
unpolarized $pp$ Drell-Yan process. The overall magnitude of the
$\cos2\phi$ dependence for $pp$ processes is consistent with, but
slightly larger than that of $p\,d$ processes. The new $pp$ data,
besides the previous $p\,d$ data, will provide further information
and constraint on the shape of the Boer-Mulders functions for
different flavors. Therefore there is the need to perform a new
extraction of Boer-Mulders function in the presence of new $pp$
data. In this work we combine the previous $p\,d$ data and the new
$pp$ data to extract Boer-Mulders functions for the proton to update
our previous results. Furthermore, in the new fit we include
$x_f$-dependent and $Q$-dependent data that have not been applied in
the previous extraction. We then apply our extracted Boer-Mulders
functions to predict the $\cos 2\phi$ asymmetries in future $pp$
experiments at J-PARC and $p\bar{p}$ experiments at PANDA and PAX.

\section{Description of $\cos2\phi$ asymmetries in terms of Boer-Mulders functions}

The angular differential cross section for unpolarized Drell-Yan
processes has the general form

\begin{eqnarray}
\frac{1}{\sigma}\frac{d\sigma}{d\Omega}&=&\frac{3}{4\pi}\frac{1}{\lambda+3}
(1+\lambda\,\cos^2\theta+\mu\,\sin2\,\theta\,\cos\phi
\nonumber\\
& & +\frac{\nu}{2}\sin^2\theta\cos2\phi).\label{cos2phi}
\end{eqnarray}
where $\theta$ and $\phi$ are, respectively, the polar angle and the
azimuthal angle of dileptons in the Collins-Soper frame~\cite{cs77}.
The coefficients $\lambda, \mu$ and $\nu$ do not depend on these
angles, and for scattering that has azimuthal symmetry their values
are $\mu = \nu = 0$.

This angular distribution has been measured in muon pair production
by pion-nucleon collisions: $\pi^-N\rightarrow\mu^+\mu^-X$, with $N$
denoting a nucleon in deuterium or tungsten, and for a $\pi^-$ beam
with energies of 140, 194, 286 GeV~\cite{na10} and 252
GeV~\cite{conway}. The experimental data show large values of $\nu$,
near 30\%.
%indicating the violation of the Lam-Tung relation $1-\lambda=2\nu$.
The most recent measurements of the angular distribution were
performed by the E866 Collaboration~\cite{e866}, in $pd$ Drell-Yan
processes at 800 GeV/c. The measured $\nu$ is about several percent,
a result which can not be explained by leading-twist collinear
factorization~\footnote{In has been shown in Ref.~\cite{Zhou:2009rp}
that the $\cos 2 \phi$ asymmetries can be explained by taking into
account the twist-three quark-gluon correlations in collinear
factorization which is consistent with the Boer-Mulders effect in
the TMD factorization approach.} in QCD. As proposed by
Boer~\cite{boer}, the non-zero $\cos 2\phi$ term can be produced by
the product of two $h_1^\p$s, each coming from one of the two
incident hadrons.

The leading order unpolarized Drell-Yan cross section expressed in
the Collins-Soper frame~\cite{cs77} is~\cite{boer}
\begin{eqnarray}
&&\frac{d\sigma(h_1h_2\rightarrow l\bar{l}X)}{d\Omega
dx_1dx_2d^2\boldsymbol{q}_T}=
\frac{\alpha^2}{3Q^2}\sum_{q,\bar{q}}\Bigg{\{}
A(y)\mathcal{F}[f_1^{\,q}f_1^{\,\bar{q}}] +B(y)\nonumber\\
&&\,\times\,\cos2\phi\mathcal{F}\left [((2\hat{\boldsymbol{h}}\cdot
\boldsymbol{p}_T\hat{\boldsymbol{h}}\cdot \boldsymbol{k}_T)
-(\boldsymbol{p}_T\cdot
\kp))\frac{h_1^{\perp\,q}h_1^{\perp\,\bar{q}}}{M_1M_2}\right
]\Bigg{\}},\nonumber\\
\label{cs}
\end{eqnarray}
where $Q^2=q^2$ and $\boldsymbol{q}_T$ are the invariance mass
square and the transverse momentum of the lepton pair. The vector
$\hat{\boldsymbol{h}}=\boldsymbol{q}_T/Q_T$. We have used the
notation
\begin{eqnarray}
\mathcal{F}[f\bar{f}]=\int d^2\boldsymbol{p}_T
d^2\kp\delta^2(\boldsymbol{p}_T+\kp-\boldsymbol{q}_T)
f(x_1,\boldsymbol{p}_T^2)\bar{f}(x_2,\kp^2).
\end{eqnarray}
The first term in Eq.~(\ref{cs}) is azimuthal independent, while the
second term has a $\cos2\phi$ azimuthal dependent term which
contributes to the asymmetry $\nu$.

In the case of the $p\,d$ and $pp$ Drell-Yan processes, the
$\cos2\phi$ asymmetry can be expressed as
\begin{eqnarray}
\nu_{pd}(x_1,x_2,q_T)&=&\frac{F_{pd}(x_1,x_2,q_T)}{M_p^2\,G_{pd}(x_1,x_2,q_T)},\\
\nu_{pp}(x_1,x_2,q_T)&=&\frac{F_{pp}(x_1,x_2,q_T)}{M_p^2\,G_{pp}(x_1,x_2,q_T)},
\end{eqnarray}
where
\begin{eqnarray}
F_{pd}(x_1,x_2,q_T)&=& 2\mathcal{F} \left
[(2\hat{\boldsymbol{h}}\cdot
\boldsymbol{p}_T\hat{\boldsymbol{h}}\cdot \boldsymbol{k}_T
-\boldsymbol{p}_T\cdot \kp)\,(e_u^2 h_1^{\perp\,u}\right.\nonumber
 \\&+&\left.
e_d^2 h_1^{\perp\,d}) (h_{1}^{\perp\,\bar{u}}
+h_{1}^{\perp\,\bar{d}})\right ]+(q
\leftrightarrow \bar{q}),\label{fpd}\\
G_{pd}(x_1,x_2,q_T)&=&\mathcal{F}\left [ (e_u^2 f_1^{u} +e_d^2 f_1^{
d}) (f_{1}^{\bar{u}} +f_{1}^{\bar{d}})\right]+(q \leftrightarrow
\bar{q}),\label{gpd} \\
F_{pp}(x_1,x_2,q_T)&=&2 \mathcal{F}\left
[(2\hat{\boldsymbol{h}}\cdot
\boldsymbol{p}_T\hat{\boldsymbol{h}}\cdot \boldsymbol{k}_T
-\boldsymbol{p}_T\cdot \kp)\, (e_u^2
h_1^{\perp\,u}h_{1}^{\perp\,\bar{u}} \right.\nonumber
\\&+&\left.
e_d^2 h_1^{\perp\,d}h_{1}^{\perp\,\bar{d}})\right ]+(q
\leftrightarrow \bar{q}),\label{fpp}\\
G_{pp}(x_1,x_2,q_T)&=& \mathcal{F}[ (e_u^2 f_1^{u}\,f_{1}^{\bar{u}}
+e_d^2 f_1^{ d}\,f_{1}^{\bar{d}})]+(q \leftrightarrow
\bar{q}).\label{gpp}
\end{eqnarray}
For distribution functions for deuteron, we have used the isospin
relation:
\begin{eqnarray}
f^{u/deuteron} \approx f^{u/p}+f^{u/n}=f^u+f^d.
\end{eqnarray}
In Ref.~\cite{blms08a} we have parameterized the transverse momentum
dependence of Boer-Mulders functions with a Gaussian form as follows
\begin{equation}
h_1^{\perp\,q}(x,\boldsymbol{p}_T^2)=h_1^{\perp\,q}(x)\frac{\exp\,(-\boldsymbol{p}_{T}^{2}/
p_{bm}^2)}{\pi p_{bm}^2}.\label{bmexp}
\end{equation}
The x dependence for $u$, $d$, $\bar{u}$ and $\bar{d}$ quarks is
parameterized, as follows \ba
h_1^{\p\,q}(x)&=&H_q\,x^c\,(1-x)\,f_1^{\,q}(x).\label{p1} \ea The
above parametrizations, with 6 parameters, have been applied to fit
$p\,d$ Drell-Yan data measured by E866/NuSea Collaboration. In the
fit the $P_T$-dependent and $x_{1/2}$-dependent $\cos2\phi$
asymmetry data were used. The fitted result was employed to predict
the $x_f$-dependent and $Q$-dependent $\cos2\phi$ asymmetries which
were compared with the corresponding data.

Recently the E866/NuSea Collaboration reports
measurements~\cite{e866pp} of the $\cos2\phi$ asymmetries on
unpolarized $pp$ Drell-Yan processes at $E_p = 800$ GeV. The new
$pp$ data, together with the previous $p\,d$ data, will provide
further information on the shape of the Boer-Mulders functions for
different flavors. In this paper we will combine the previous $p\,d$
data and the new $pp$ data in the fit. Further more we will include
$x_f$-dependent and $Q$-dependent data in our fit. To do this we
will parameterize the Boer-Mulders functions as in
Eqs.~(\ref{bmexp}) and (\ref{p1}), but changing the form slightly.

\begin{figure}
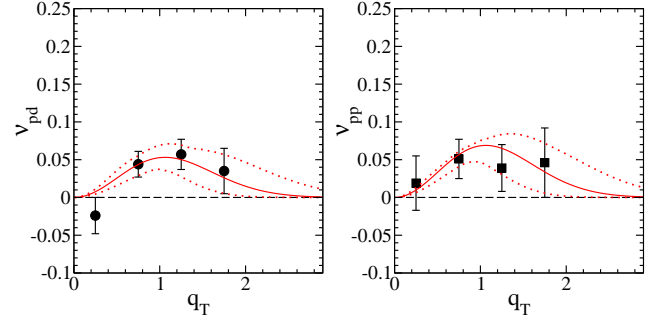

\begin{center}
\scalebox{0.24}{\includegraphics*[0pt,24pt][496pt,529pt]{pd-qt-err.eps}}
\scalebox{0.24}{\includegraphics*[0pt,24pt][496pt,520pt]{pp-qt-err.eps}}
\caption{\small The $q_T$-dependent $\cos 2 \phi$ asymmetries for
unpolarized $p\,d$ (left) and $pp$ (right) Drell-Yan process
calculated from our fitted results. The region between the two
dotted line corresponds to the uncertainty of the parameters. Data
are from the FNAL E866/NuSea experiments. Here we exclude data with
$p_T
>2$ GeV.}\label{pdppqt}
\end{center}
\end{figure}

\begin{figure}
\begin{center}
\scalebox{0.25}{\includegraphics*[1pt,24pt][486pt,518pt]{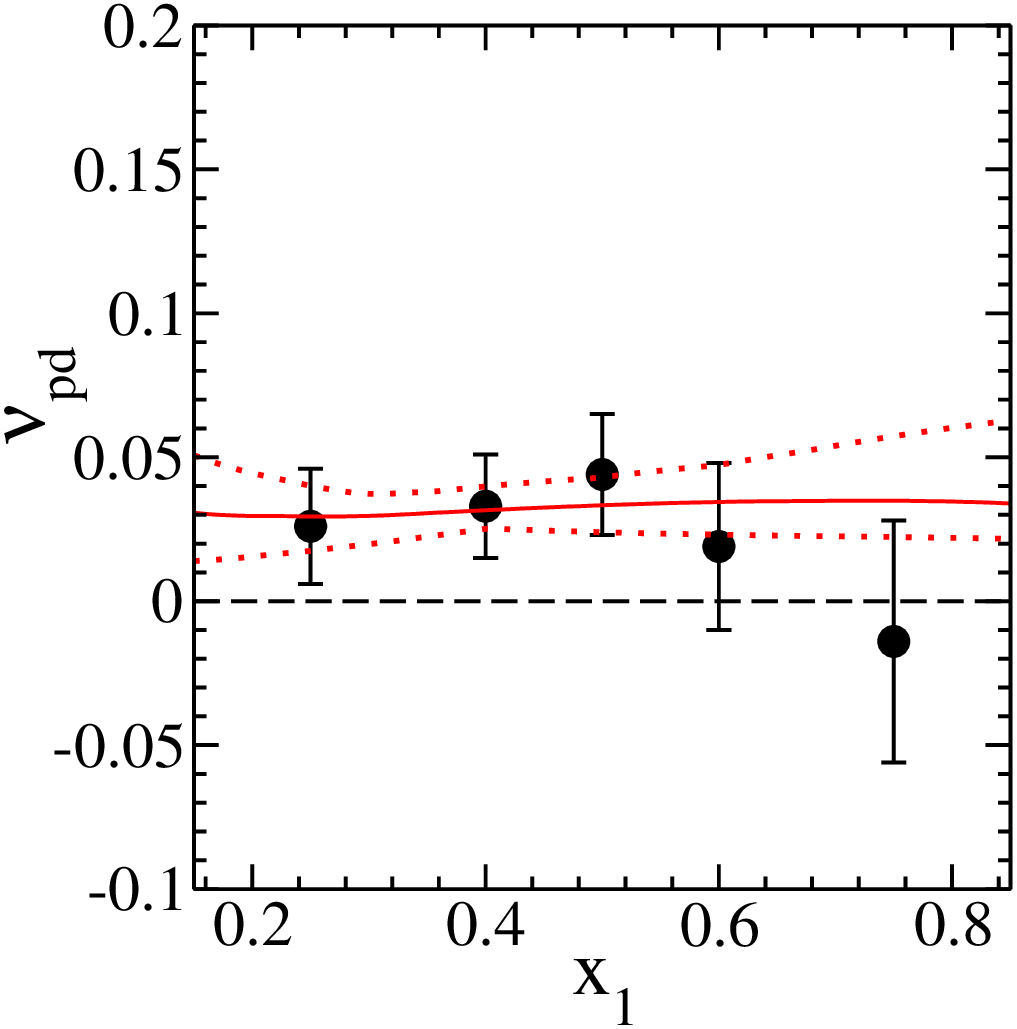}}
\scalebox{0.25}{\includegraphics*[1pt,24pt][487pt,518pt]{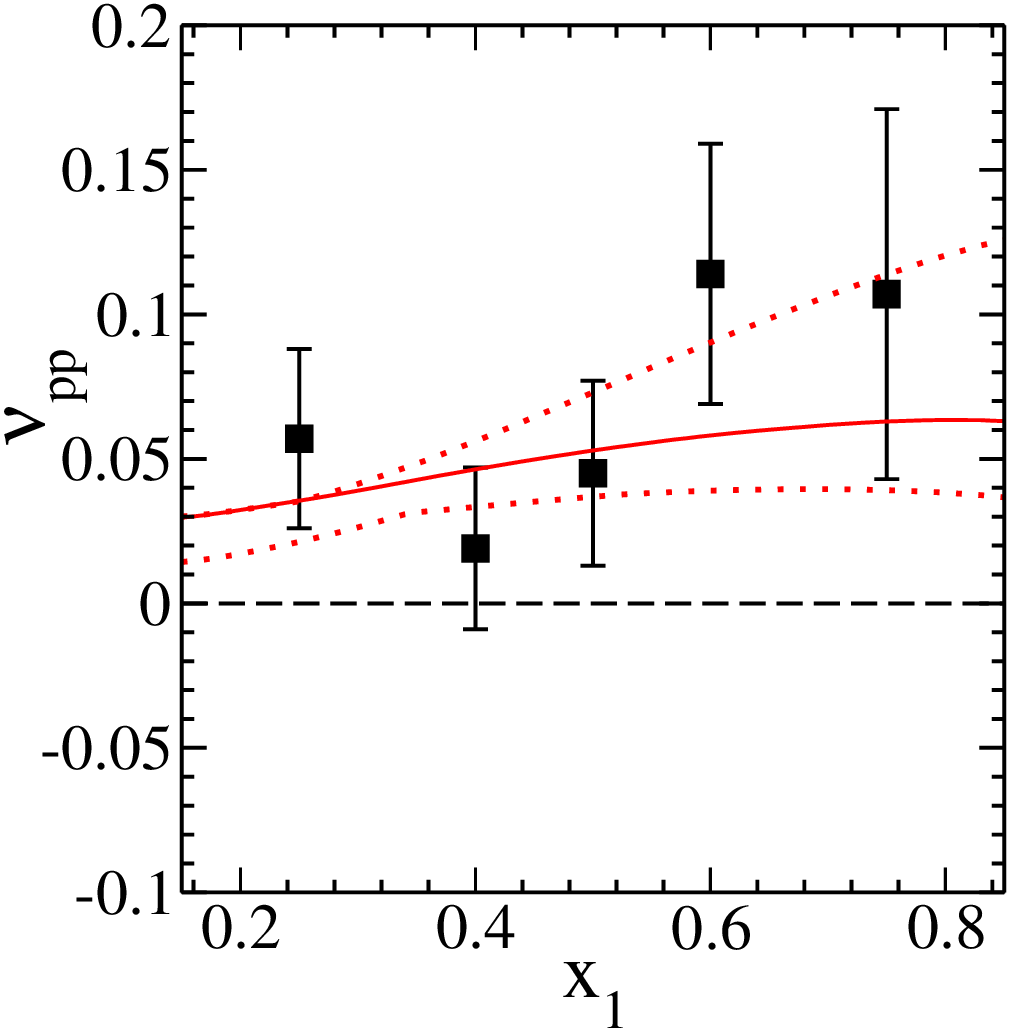}}
\caption{\small The $x_1$-dependent $\cos 2 \phi$ asymmetries for
unpolarized $p\,d$ (left) and $pp$ (right) Drell-Yan process
calculated from our fitted results. The region between the two
dotted line corresponds to the uncertainty of the parameters. Data
are from the FNAL E866/NuSea collaboration.}\label{pdppx1}
\end{center}
\end{figure}

\begin{figure}
\begin{center}
\scalebox{0.25}{\includegraphics*[1pt,24pt][487pt,518pt]{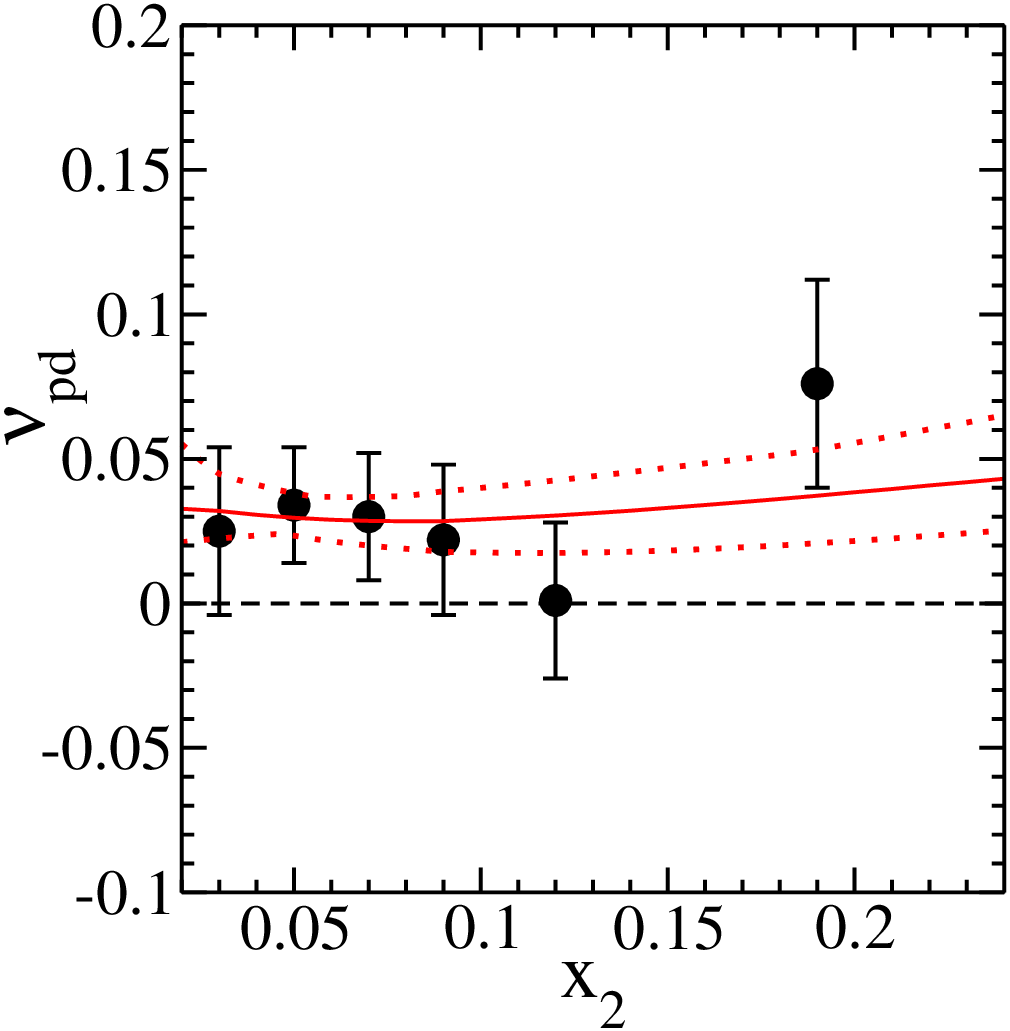}}
\scalebox{0.25}{\includegraphics*[1pt,24pt][487pt,518pt]{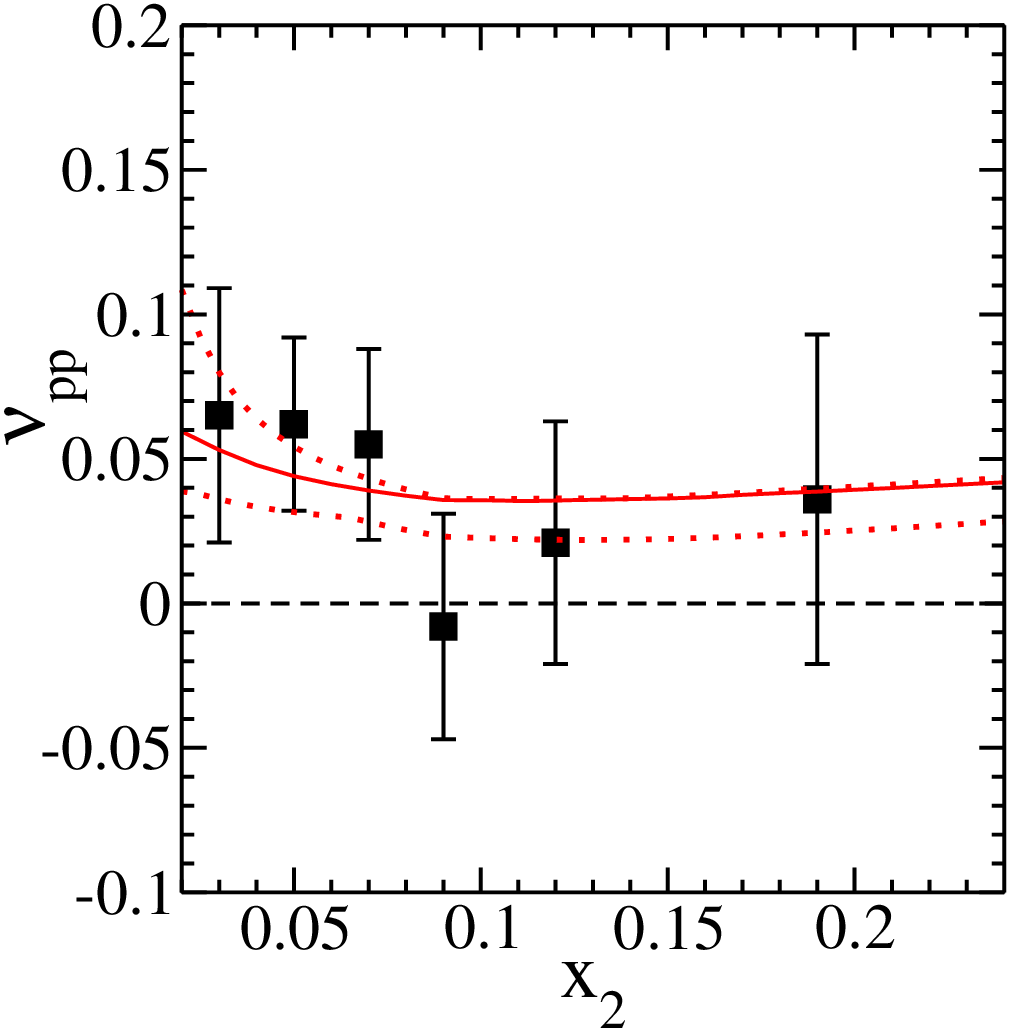}}
\caption{\small The $x_2$-dependent $\cos 2 \phi$ asymmetries for
unpolarized $p\,d$ (left) and $pp$ (right) Drell-Yan process
calculated from our fitted results. The region between the two
dotted line corresponds to the uncertainty of the parameters. Data
are from the FNAL E866/NuSea collaboration.}\label{pdppx2}
\end{center}
\end{figure}

In our previous fit we modeled the $x$-dependent behavior of
$h_1^{\p\,q}(x,\boldsymbol{p}_T^2)$ at small $x$ as $x^c$ compared
with $f_1^{\,q}(x)$, and we assumed the value of $c$ to be flavor
independent, as shown in (\ref{p1}). Now with more data available,
we are able to release this constraint to replace $c$ as $c_q$,
depending on flavor. Secondly we model the large $x$-dependence of
the Boer-Mulders functions by $(1-x)^b$, different from our previous
fit in which the large $x$ dependence is $1-x$. Therefore we have
the new parametrizations for $h_1^{\p \,q} $ ($q=u, d ,\bar{u}$ and
$\bar{d}$) as follows: \ba h_1^{\p\,
q}(x)&=&H_{q}\,x^{c^{q}}\,(1-x)^b\,f_1^{\,q}(x),\label{hq_new} \ea

The first $\boldsymbol{p}_T^2$-moment of Boer-Mulders function is
defined as
\begin{equation}
h_1^{\p\,(1)\,q}(x)=\int \textrm{d}^2 \, \boldsymbol{p}_T \,
\frac{\boldsymbol{p}_T^2}{2M^2}\,h_1^{\p}(x,\boldsymbol{p}_T^2)
\end{equation}
From Eqs.~(\ref{bmexp}) and (\ref{hq_new}) one can calculate
$h_1^{\p\,(1)\,q}(x)$ from our parametrization as
\begin{equation}
h_1^{\p\,(1)\,q}(x)=\frac{p_{bm}^2}{2M^2}h_1^{\p \,q}(x)
\end{equation}

\begin{figure}
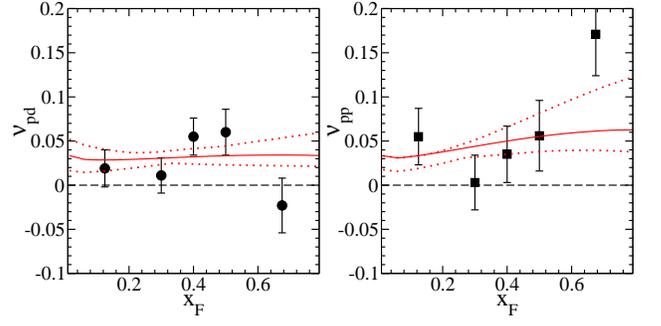

\begin{center}
\scalebox{0.24}{\includegraphics*[1pt,24pt][486pt,518pt]{pd-xf-err.eps}}
\scalebox{0.24}{\includegraphics*[1pt,24pt][486pt,518pt]{pp-xf-err.eps}}
\caption{\small The $x_F$-dependent $\cos 2 \phi$ asymmetries for
unpolarized $p\,d$ (left) and $pp$ (right) Drell-Yan process
calculated from our fitted results. The region between the two
dotted line corresponds to the uncertainty of the parameters. Data
are from the FNAL E866/NuSea collaboration.}\label{pdppxf}
\end{center}
\end{figure}

\begin{figure}
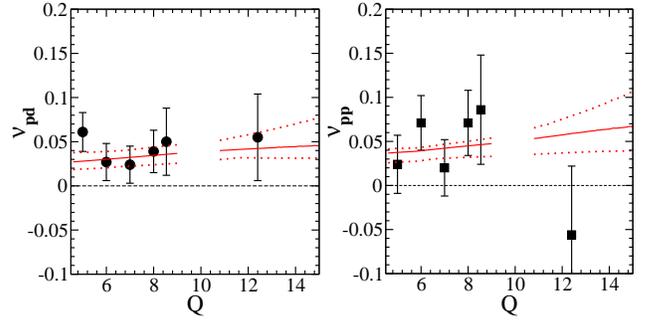

\begin{center}
\scalebox{0.24}{\includegraphics*[1pt,23pt][486pt,518pt]{pd-q-err.eps}}
\scalebox{0.24}{\includegraphics*[1pt,23pt][486pt,518pt]{pp-q-err.eps}}
\caption{\small The $Q$-dependent $\cos 2 \phi$ asymmetries for
unpolarized $p\,d$ (left) and $pp$ (right) Drell-Yan process
calculated from our fitted results. The region between the two
dotted line corresponds to the uncertainty of the parameters. Data
are from the FNAL E866/NuSea collaboration.}\label{pdppq}
\end{center}
\end{figure}

With the Gaussian form for the $\boldsymbol{p}_T$ dependence of
Boer-Mulders functions and the unpolarized TMD distribution
\begin{equation}
f_1^{\,q}(x,\boldsymbol{p}_T^2)=f_1^{\,q}(x)\frac{1}{\pi
p_{un}^2}\exp\left(\frac{\boldsymbol{p}_T^2}{p_{un}^2}\right),
\end{equation}
 the transverse momentum integrations in Eqs.~(\ref{fpd}) --
(\ref{gpp}) can be deconvoluted and the results are:
\begin{eqnarray}
F_{pd}(x_1,x_2,q_T)&=&F_{pd}(x_1,x_2) \frac{q_T^2}{36\pi
p_{bm}^2}\exp\left(-\frac{q_T^2}{2p_{bm}^2}\right),\label{fpd1}\\
%%%%%%%%%%%%%%%%%%%%%%%%%%%%%%%%%%%%%%%%%%%%%%%%%%%%%%%%%%%%%%%
F_{pp}(x_1,x_2,q_T)&=&F_{pp}(x_1,x_2)\frac{q_T^2}{36\pi
p_{bm}^2}\exp\left(-\frac{q_T^2}{2p_{bm}}\right),\label{fpp1}\\
%%%%%%%%%%%%%%%%%%%%%%%%%%%%%%%%%%%%%%%%%%%%%%%%%%%%%%%%%%%%%%%%%%%%
G_{pd}(x_1,x_2,q_T)&=& G_{pd}(x_1,x_2)\frac{1}{18\pi p_{un}^2}\exp
\left(-\frac{q_T^2}{2p_{un}^2}\right),\label{gpd1} \\
%%%%%%%%%%%%%%%%%%%%%%%%%%%%%%%%%%%%%%%%%%%%%%%%%%%%%%%%%%%%%
G_{pp}(x_1,x_2,q_T)&=& G_{pp}(x_1,x_2)\frac{1}{18\pi p_{un}^2}
\exp\left(-\frac{q_T^2}{2p_{un}^2}\right). \label{gpp1}
\end{eqnarray}
where the $x_{1/2}$ dependent parts are
\begin{eqnarray}
&&F_{pd}(x_1,x_2)= \,\big{(}\,4\,h_1^{\perp\,u}\,(x_1)
+\,h_1^{\perp\,d}(x_1)\,\big{)}\nonumber\\ &\times &
\big{(}\,h_{1}^{\perp\,\bar{u}}(x_2)\,+\,
  h_{1}^{\perp\,\bar{d}}(x_2)\,\big{)} + (q \, \rightarrow \,\bar{q})
\nonumber\\ &=&
(1-x_1)^b\,(1-x_2)^b(4H_1\,x_1^{c_u}\,f_1^u(x_1)\,x_2^{c_{\bar{u}}}\,f_1^{\bar{u}}(x_2)
\nonumber\\
&+&H_2\,x_1^{c_d}\,f_1^d(x_1)\,x_2^{c_{\bar{d}}}f_1^{\bar{d}}(x_2)\,+\,
4H_3\,x_1^{c_u} \,f_1^u(x_1)\,x_2^{c_{\bar{d}}}\,f_1^{\bar{d}}(x_2)
\nonumber\\
&+&(H_1H_2/H_3)\,x_1^{c_d}\,f_1^d(x_1)\,x_2^{c_{\bar{d}}}\,
f_1^{\bar{u}}(x_2))+ (q \, \rightarrow \,\bar{q}),\label{fpd2}\\
%%%%%%%%%%%%%%%%%%%%%%%%%%%%%%%%%%%%%%%%%%%%%%%%%%%%%%%%%%%%%%%%%%%%%%%%%%%%%%%%%%%
&&F_{pp}(x_1,x_2)= 4\,h_1^{\perp\,u}\,(x_1) h_{1}^{\perp\,\bar{u}}
(x_2) + h_1^{\perp\,d}(x_1) h_{1}^{\perp\,\bar{d}}(x_2)+ (q
\rightarrow \bar{q})\nonumber\\
&=&(1-x_1)^b\,(1-x_2)^b(4H_1\,x_1^{c_u}\,f_1^u(x_1)\,x_2^{c_{\bar{u}}}
\,f_1^{\bar{u}}(x_2) \nonumber\\
&+&H_2\,x_1^{c_d}\,f_1^d(x_1)\,x_2^{c_{\bar{d}}}\,f_2^{\bar{d}}(x_2))
+ (q \, \rightarrow \,\bar{q}),\label{fpp2}\\
%%%%%%%%%%%%%%%%%%%%%%%%%%%%%%%%%%%%%%%%%%%%%%%%%
& &G_{pd}(x_1,x_2)= \big{(}4 f_1^{u}\,(x_1)\, +\, f_1^{
d}\,(x_1)\,\big{)} \nonumber
 \\&\times&
\big{(}f_{1}^{\bar{u}}\,(x_2)\,
+f_{1}^{\bar{d}}\,(x_2)\,\big{)}+ (q \, \rightarrow \,\bar{q}),\label{gpd2} \\
 %%%%%%%%%%%%%%%%%%%%%%%%%%%%%%%%%%%%%%%%%%%%%%%%%%%%%%%%%%%%%
& & G_{pp}(x_1,x_2)= 4 f_1^{u}(x_1)\,f_{1}^{\bar{u}}(x_2)\, +\,
f_1^{ d}(x_1)\,f_{1}^{\bar{d}}(x_2) + (q \rightarrow
\bar{q}), \nonumber \label{gpp2} \\
\end{eqnarray}
where $H_1=H_u\,H_{\bar{u}}$, $H_2 = H_d\,H_{\bar{d}}$, $H_3 =
H_u\,H_{\bar{d}}$ and $H_1H_2/H_3 = H_{\bar{u}}\,H_d$. Since
$H_{u}$, $H_{d}$, $H_{\bar{u}}$ and $H_{\bar{u}}$ always appear as
 products of two of them, we will apply $H_1$, $H_2$ and $H_3$ as
the parameters in the fit. Therefore the actual number of free
parameters is reduced to 9.

The $q_T$-, $x_1$- and $x_2$-dependent $\cos2\phi$ asymmetries in
unpolarized $p\,d$ and $pp$ Drell-Yan processes can then be
expressed as
\begin{eqnarray}
\nu_{NN}(q_T)&=&\frac{\int d x_1\int d x_2 \,F_{NN}(x_1,x_2,q_T)}
{M_p^2\,\int d x_1\int d x_2\,G_{NN}(x_1,x_2,q_T)},\label{nuqt}\\
%%%%%%%%%%%%%%%%%%%%%%%%%%%%%%%%%%%%%%%%%%%%%%%%%%%%%%%%%%%%%%%%%%%%%%
\nu_{NN}(x_1)&=&\frac{\int d x_2\int d q_T^2 \,F_{NN}(x_1,x_2,q_T)}
{M_p^2\,\int d x_2\int d q_T^2\,G_{NN}(x_1,x_2,q_T)},\label{nux1}\\
%%%%%%%%%%%%%%%%%%%%%%%%%%%%%%%%%%%%%%%%%%%%%%%%%%%%%%%%%%%%%%%%%%%%%%
\nu_{NN}(x_2)&=&\frac{\int d x_1\int d q_T^2
\,F_{NN}(x_1,x_2,q_T)}{M_p^2\,\int d x_1\int d
q_T^2\,G_{NN}(x_1,x_2,q_T)},\label{nux2}
\end{eqnarray}
where the subscript $NN$ denotes $p\,d$ and $pp$.

\begin{figure}
\begin{center}
\scalebox{0.18}{\includegraphics*[8pt,51pt][659pt,617pt]{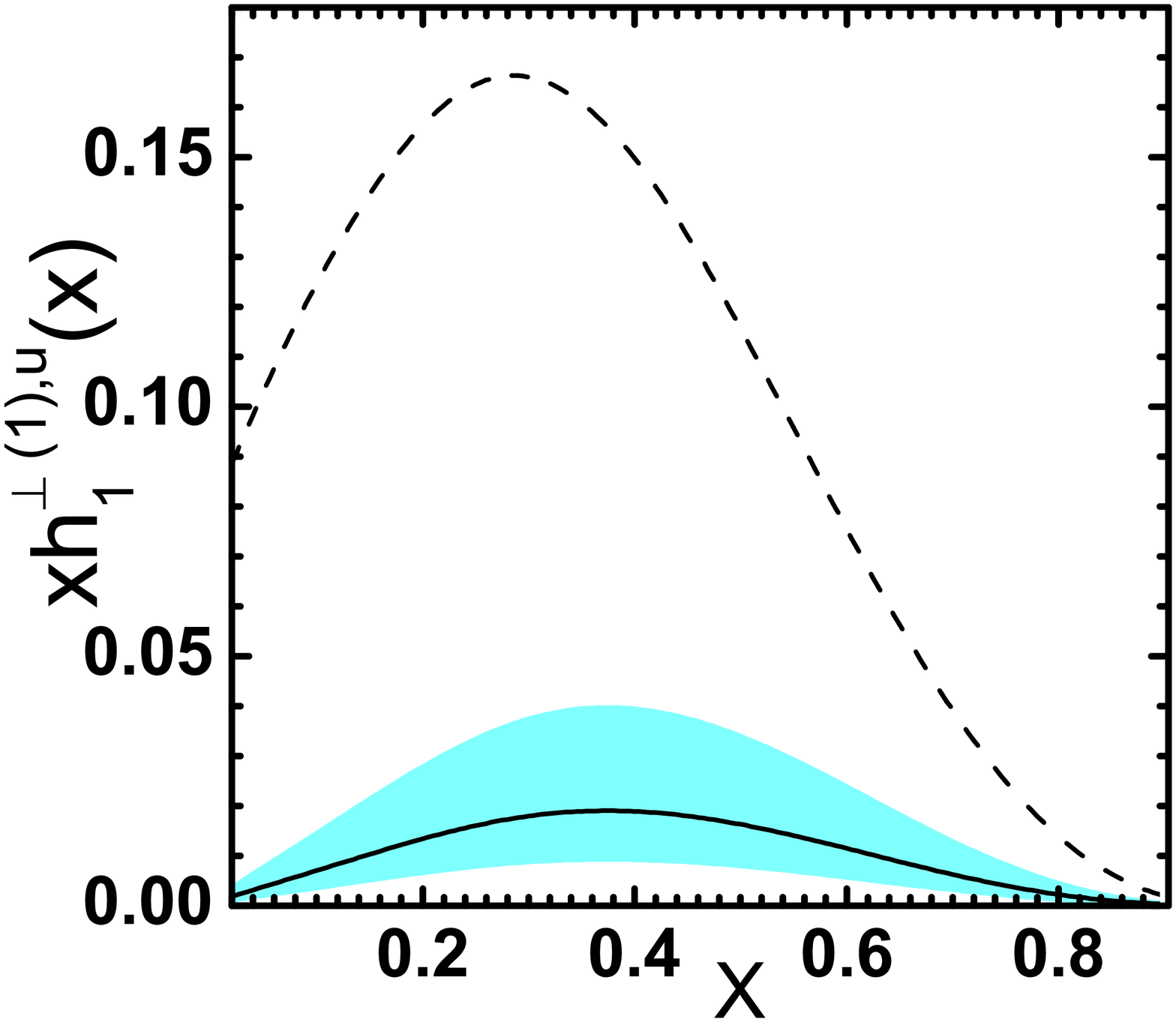}}
\scalebox{0.18}{\includegraphics*[8pt,51pt][659pt,617pt]{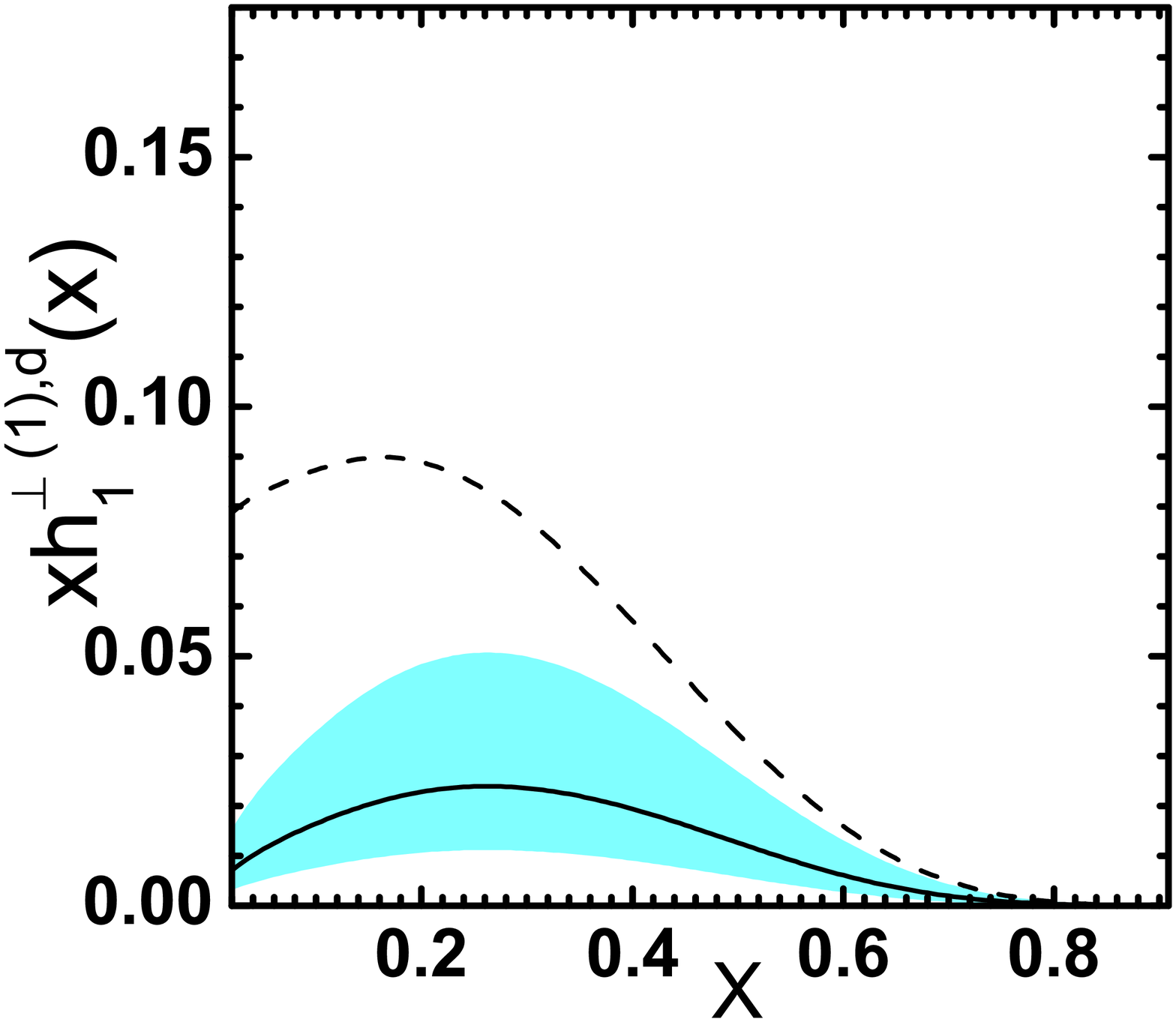}}\\
\scalebox{0.18}{\includegraphics*[8pt,51pt][659pt,617pt]{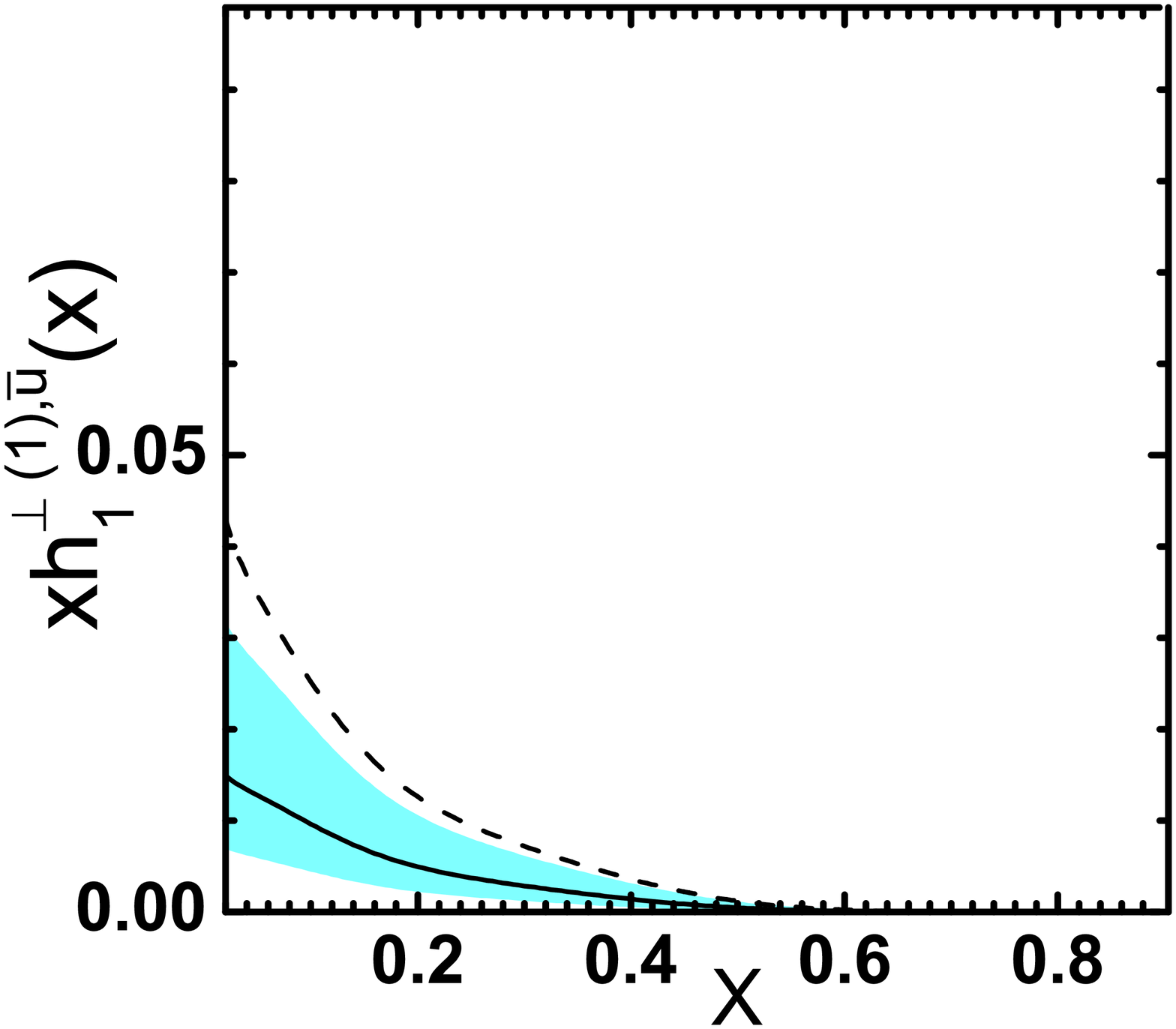}}
\scalebox{0.18}{\includegraphics*[8pt,51pt][659pt,617pt]{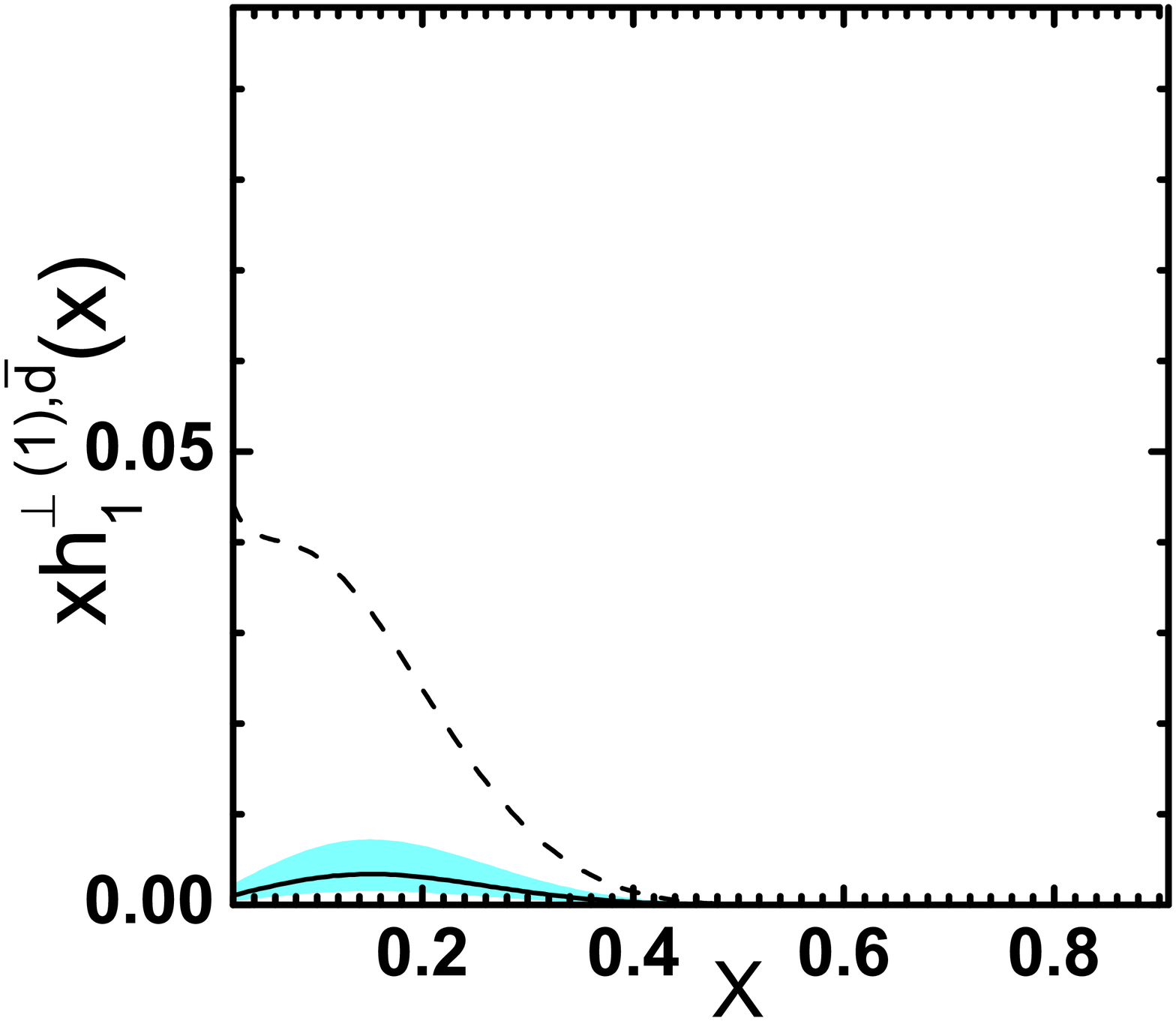}}
%\scalebox{0.18}{\includegraphics*[26pt,44pt][660pt,529pt]{Graph13-a.eps}}
%\scalebox{0.33}{\includegraphics[80pt,70pt][490pt,460pt]{pdx2.eps}}
\caption{\small The first $\boldsymbol{p}_T^2$-moments of
Boer-Mulders functions for $u$, $d$, $\bar{u}$ and $\bar{d}$ quarks
for $Q^2=1$ GeV$^2$ by solid lines, the shadows depict the %%correction: 2 GeV^2 should be 1 GeV^2
variation ranges of $x\,h_1^{\p\,(1)\,q}(x)$ allowed by the
positivity bound. The dashed lines show $\frac{\langle p_T
\rangle_{un} }{2M} x\,f_{\,1}^{\,q}\,(x\,)$.
.}\label{fig_bm_moments}
\end{center}
\end{figure}

One can also express the cross-section of the Drell-Yan process,
depending on Feynman $x_F$ and the mass of the lepton pair $Q$ as
\begin{eqnarray}
\frac{d\sigma}{dx_F dQ^2 d^2\boldsymbol{q}_T} =
\frac{1}{s\sqrt{x_F^2+Q^2/s}}\frac{d\sigma}{dx_1 dx_2
d^2\boldsymbol{q}_T},
\end{eqnarray}
with
\begin{eqnarray}
x_{1/2}=\frac{\pm x_F+\sqrt{x_F^2+Q^2/s}}{2},~~~ Q^2=x_1\,x_2\,s.
\end{eqnarray}
Therefore the $x_F$- and $Q$-dependent $\cos2\phi$ asymmetries can
be expressed as
\begin{eqnarray}
\nu_{NN}(x_F)&=&\frac{\int dQ^2\int dq_T^2
\,\frac{F_{NN}(x_1,x_2,q_T)}{\sqrt{x_F^2+Q^2/s}}}
{M_p^2\,\int dQ^2\int d q_T^2\, \frac{G_{NN}(x_1,x_2,q_T)}{\sqrt{x_F^2+Q^2/s}}},\label{nuxf}\\
%%%%%%%%%%%%%%%%%%%%%%%%%%%%%%%%%%%%%%%%%%%%%%%%%%%%%%%%%%%%%%%%%%%%%%%%%%%%%%%%
\nu_{NN}(Q)&=&\frac{\int dx_F\int dq_T^2
\,\frac{F_{pd}(x_1,x_2,q_T)}{\sqrt{x_F^2+Q^2/s}}} {M_p^2\,\int
x_F\int q_T^2
\,\frac{G_{NN}(x_1,x_2,q_T)}{\sqrt{x_F^2+Q^2/s}}}.\label{nuq}
\end{eqnarray}

\section{fitting Boer-Mulders functions to the unpolarized E866/NuSea $p\,d$ and $pp$ data}

The E866/NuSea Collaboration measured the $\cos 2\phi$ asymmetryies
$\nu_{pd}$ and $\nu_{pd}$ vs $Q_T$, $x_1$, $x_2$, $x_F$ and
$m_{\mu\mu}$ in the following kinematical region:
\begin{eqnarray}
&& 4.5\,\textrm{GeV}<Q
<9\,\textrm{GeV}~~\textrm{and}~~10.7\,\textrm{GeV}<Q<15\,\textrm{GeV},\nonumber\\
&&q_T<4~\textrm{GeV},~~~0.15< x_1 <0.85,~~~0.02<x_2<0.24.
\nonumber\label{e866cut}
\end{eqnarray}

In the following we apply the theoretical expressions (\ref{nuqt}) -
(\ref{nux2}), (\ref{nuxf}) and (\ref{nuq}) to fit the unpolarized
$p\,d$ and $pp$ Drell-Yan $\cos 2\phi$ asymmetry
data~\cite{e866,e866pp}. The Boer-Mulders effect to the $\cos 2\phi$
asymmetry is supposed to apply in the region where $q_T$ is not
large. At large $q_T$, the higher order perturbative QCD
contributions~\cite{Boer:2006eq,Berger:2007si} might be important.
Therefore we exclude the data with $q_T > 2$~GeV in our fit. For the
parton distribution $f_1^q(x)$ we adopt the MSTW2008 LO
set~\cite{mstw2008}. We choose the Gaussian width for
$f_1^{q}(x,\boldsymbol{p}_T^2)$  as $p_{un}^2=0.25\,
\textrm{GeV}^2$, following the value given in
Refs.~\cite{Anselmino:2005nn,Anselmino:2005ea}. The best fit results
and the errors for the parameters are as follows:
\begin{eqnarray}
       H_1 &=&0.62^{+0.52}_{-0.29},~~
      H_2= 1.45^{+1.30}_{-1.12},~~
      H_3 = 0.61^{+0.50}_{-0.55},~~ \nonumber\\
       c_{u}&=&  0.63^{+0.53}_{-0.21},~~~
       c_{d} =  0.47^{+0.36}_{-0.39},~~~
       c_{\bar{u}}=  \,0.07^{+0.06}_{-0.05},~~\label{fitres}\\
       c_{\bar{d}}&=& 0.75^{+0.72}_{-0.52},~~~
       b_0 = 0.17^{+0.15}_{-0.14},~
       p_{bm}^2=0.173^{+0.027}_{-0.033}.\nonumber
\end{eqnarray}
 The $\chi^2$ of this fit is 35.95 for 52 data points, resulting
$\chi^2/d.o.f=0.84$. In
Figs.~\ref{pdppqt},~\ref{pdppx1},~\ref{pdppx2},~\ref{pdppxf} and
\ref{pdppq} we show the $q_T$-, $x_1$-, $x_2$-, $x_F$- and
$Q$-dependent $\cos 2 \phi$ asymmetries for unpolarized $p\,d$ and
$pp$ Drell-Yan process calculated from our fitted results and
compare them with FNAL E866/NuSea data. The solid lines show the
best fit results, and the regions between the two dotted lines
correspond to the uncertainty of the parameter. In Fig.~\ref{pdppqt}
we also show the predictions for $p_T>2 GeV$ region from the
Boer-Mulders effect.

The possible range of coefficients $H_q$ are obtained from the
values of $H_1$, $H_2$ and $H_3$, by employing the positivity
bound~\cite{bacchetta00} for $h_1^{\p\,q} (x, \boldsymbol{p}_T^2)$
for the entire $x$ and $\boldsymbol{p}_T$ regions:
\begin{equation}
\frac{|p_{T} h^{\p\,q}_1(x,\boldsymbol{p}
 _{T}^2)|}{M} \le f_1^{\,q}(x,\boldsymbol{p}_{T}^2).\label{posbound}
\end{equation}
We have
\begin{eqnarray}
H_u &=& 0.59^{\,+\,0.64}_{\,-\,0.31},~~~ H_d =1.37^{\,+\,1.53}_{\,-\,0.72},\nonumber\\
H_{\bar{u}} &=& 1.10^{\,+\,1.21}_{\,-\,0.57},~~~ H_{\bar{d}} =1.08
^{\,+\,1.18}_{\,-\,0.56}.\label{hq_para}
\end{eqnarray}
The upper and lower limits for $H_q$ are determined by the
positivity bound for $h_1^{\p\,q} (x, \boldsymbol{p}_T^2)$. The
central value for $H_q$ shown above is obtained from the geometric
mean values of the upper and lower limits for $H_q$:
$H_q^{\,\textrm{cen}}=\sqrt{H_q^{\,\textrm{max}}H_q^{\,\textrm{min}}}$.
In our previous work \cite{blms08a} the variation range of $H_q$
allowed by the positivity bound was described by the coefficient
$\omega$, namely, that the substitution $H_q \to \omega H_q$ for
$q=u$, $d$ and $H_q\to \frac{1}{\omega} H_q$ for $q = \bar{u}$,
$\bar{d}$ will not change the result. In our new fit presented here,
the range of $\omega$ is $0.48 < \omega < 2.1$, and central values
for $H_q$ correspond to $\omega=1$.

The positivity bound given in (\ref{posbound}) implies
\begin{equation}
h_1^{\p (1)\,q}(x) \le \frac{\langle p_T \rangle_{un} }{2M}
f_1^q(x).
\end{equation}
In Fig.~\ref{fig_bm_moments} we show the first
$\boldsymbol{p}_T^2$-moments of the Boer-Mulders functions
$x\,h_1^{\p\,(1)\,q}(x)$ for $u$, $d$, $\bar{u}$ and $\bar{d}$
quarks for $Q^2=1$ GeV$^2$ by solid lines, the shadows depict the %%correction: 2 GeV^2 should be 1 GeV^2
variation ranges of $x\,h_1^{\p\,(1)\,q}(x)$ allowed by the
positivity bound. The dashed lines show $\frac{\langle p_T
\rangle_{un} }{2M} x\,f_{\,1}^{\,q}\,(x\,)$.

\begin{figure}
\begin{center}
\scalebox{0.32}{\includegraphics*[17pt,35pt][522pt,516pt]{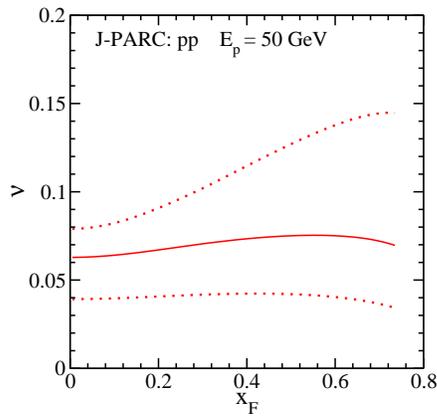}}
\caption{\small The $\cos 2\phi$ asymmetries vs $x_F$ at J-PARC with
proton beam energy $E_p = 50 $ GeV. The region between the two
dotted line corresponds to the uncertainty of the
parameters.}\label{jparc-fig}
\end{center}
\end{figure}

Several comments on our fit are in order. First, since we don't know
the scale dependence of Boer-Mulders functions, we assumed that it
has the same behavior as that of the unpolarized distribution
$f_1^q$. Recently there has been growing interest on performing
next-to-leading order
analysis~\cite{Kang:2008ey,Zhou:2008mz,Vogelsang:2009pj,Braun:2009mi,Ratcliffe:2009pp}
of single-spin asymmetries in Semi-inclusive deeply inelastic
scattering and Drell-Yan processes, especially on the Sivers
asymmetry. Those studies can provide information on the evolution of
the Sivers function. It is also interesting to study the evolution
of the Boer-Mulders function. However, to investigate the scale
dependence behavior of the Boer-Mulders function and its impacts on
the $\cos2\phi$ asymmetries is out of the scope of this paper.
Secondly, we considered the Boer-Mulders effect as the dominant
source for the $\cos 2\phi$ asymmetries, which is reasonable in the
region $q_T^2 \ll Q^2$. Since the data in E866/NuSea covers the
kinematics regime $4.5\,\textrm{GeV}<Q
<9\,\textrm{GeV}~~\textrm{and}~~10.7\,\textrm{GeV}<Q<15\,\textrm{GeV}$,
one expects that the Boer-Mulders effect dominates in the region
$q_T<2$~GeV. At $q_T>3$~GeV, the higher order perturbative QCD
contributions~\cite{Boer:2006eq,Berger:2007si} might be important.
This can be seen from Fig.~\ref{pdppqt}, which indicates that the
predicted asymmetries at large $q_T$ are small, while the size of
the data in that region is substantial~\cite{e866,e866pp}.
Thirdly, the $\chi^2/d.o.f$ in the fit presented in this work is a
little bigger than that in the previous fit shown in
Ref.~\cite{blms08a}. This is because in Ref.~\cite{blms08a} we only
included 16 data points from the $p\,d$ process, while now we have
52 data points in our new fit. And among those data we include $Q$-
and $x_F$-dependent data also. We have checked that the two data
points in the large $x_F$ region give rise to substantial
contributions ($20\%$) to the total $\chi^2$ in our best fit.

\section{predictions for future $pp$ and $p\bar{p}$ experiments}

We will then apply our extracted Boer-Mulders in previous section to
predict the $\cos 2\phi$ asymmetries in future $pp$ and $p\bar{p}$
experiments.

Drell-Yan process has been proposed at J-PARC~\cite{jparc} by pp
scattering with proton beam energy $E_p = 50$ GeV. with this lower
beam energy than that at E866/NuSea, the measurement of $\cos 2
\phi$ asymmetry at J-PARC will provide complementary information on
Boer-Mulders functions in a different kinematical region. We
estimate the $x_F$- dependent $\cos 2\phi$ asymmetry at J-PARC by
imposing the cuts $0 \leq q_T \leq 1$ GeV and $4 \leq Q \leq 5$ GeV
from the fitted results in Eq.~(\ref{fitres}) directly, as shown in
Fig.~\ref{jparc-fig}. The solid curve shows the result from the best
fitted values, while the region between the dotted lines correspond
to the parameters uncertainty.

\begin{figure}
\begin{center}
\scalebox{0.18}{\includegraphics[33pt,10pt][675pt,620pt]{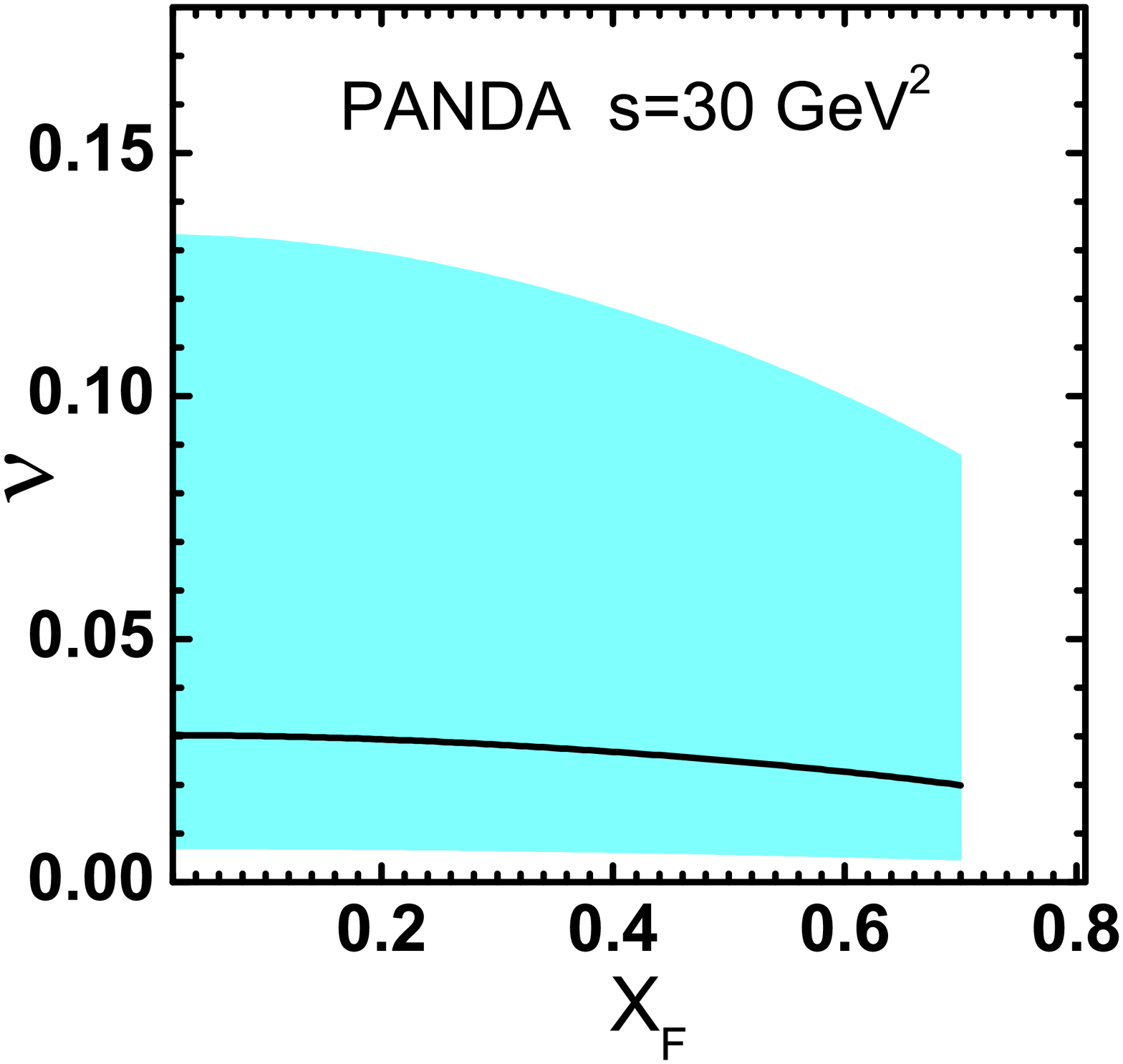}}
\scalebox{0.18}{\includegraphics[33pt,10pt][675pt,620pt]{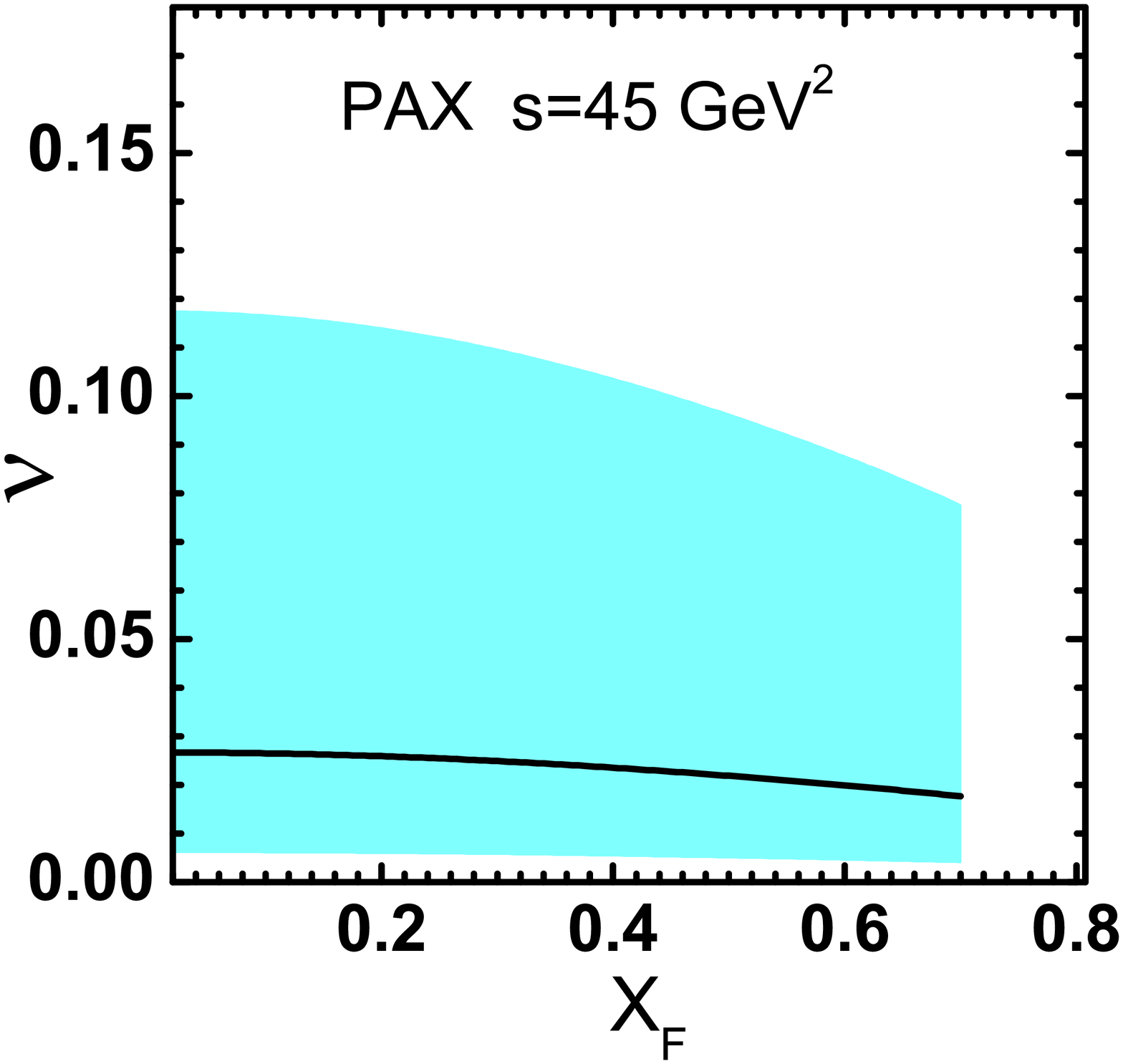}}
\caption{\small The $\cos 2\phi$ asymmetries vs $x_F$ at PANDA for
$s=30$ and at PAX for $s=45$ GeV$^2$, respectively. The bands
correspond to all possible values of $\omega$ in the allowed range
$0.48<\omega<2.1$.}\label{panda-fig}
\end{center}
\end{figure}
The GSI-PANDA experiment~\cite{PANDA} will run the Drell-Yan process
with an unpolarized antiproton beam colliding with an unpolarized
proton target at $s=30$ GeV$^2$. The PAX~\cite{PAX} experiment might
also perform unpolarized $p\bar{p}$ Drell-Yan process at the energy
$s=45$ GeV$^2$ in the fixed target mode. The kinematical cuts we
apply in the estimation of the asymmetry for those experiments are
$2\leq Q \leq 3 $ GeV and $0 \leq q_T \leq 0.4$ GeV. To estimate the
asymmetries in $p\bar{p}$ processes we need to use the $H_q$ results
given in ~(\ref{hq_para}), rather than to the results for $H_i$
($i=1,2,3$) in (\ref{fitres}). In Fig.~\ref{panda-fig} we show the
predicted $\cos 2\phi$ asymmetries $\nu$ at PANDA for $s=30$
GeV$^2$, and PAX for $s=45$ GeV$^2$, as a function of $x_F$. The
solid lines are calculated from the central value of $H_q$ given in
Eq.~(\ref{hq_para}), and the bands correspond to all possible values
of $\omega$ in the allowed range $0.48<\omega<2.1$. The asymmetry
calculated from the central value of $H_q$ in $p \bar{p}$ process is
smaller than that in $pp$ case. Therefore the measurement on the
$\cos 2 \phi$ asymmetry at PANDA and PAX might provide valuable
examination on our extraction of Boer-Mulders functions.

\section{summary}

We have parameterized the Boer-Mulders functions for the proton by
employing a Gaussian form for their transverse momentum dependence.
We then fitted our parameterizations to both the previous measured
$p\,d$ data and recent $pp$ data on the unpolarized Drell-Yan $\cos
2\phi$ dilepton asymmetries from the E866/NuSea Collaboration. The
basic assumption in the fit is that the $\cos 2\phi$ asymmetry is
contributed by the product of two Boer-Mulders functions and that
other contributions such as perturbative QCD effects can be ignored
in the region $q_T^2\ll Q^2 $. In our fit we included not only the
$q_T$-, $x_1$- and $x_2$-dependent asymmetry data, but also $Q$- and
$x_F$-dependent data. We applied our extracted Boer-Mulders to
predict the $\cos 2\phi$ asymmetries in future $pp$
 experiment at J-PARC and $p\bar{p}$ experiment at PANDA
and PAX. We found a smaller asymmetry for $p \bar{p}$ processes
compared with that for $pp$ processes, which might serve as a test
on our extraction of Boer-Mulders functions.

{\bf Acknowledgements.} We would like to thank B.-Q. Ma and B. Zhang
for useful discussions. This work is supported by the PBCT project
No. ACT-028 ``Center of Subatomic Physics" and FONDECYT (Chile)
Project No. 11090085.

\end{document}